# Pneumonia after bacterial or viral infection preceded or followed by radiation exposure - a reanalysis of older radiobiological data and implications for low dose radiotherapy for COVID-19 pneumonia

Mark P Little[a,1], Wei Zhang[b], Roy van Dusen[c], Nobuyuki Hamada[d]

[a]Radiation Epidemiology Branch, Division of Cancer Epidemiology and Genetics, National Cancer Institute, NIH, DHHS, 9609 Medical Center Drive, Rockville, MD 20892-9778

[b]Centre for Radiation, Chemical and Environmental Hazards, Public Health England, Chilton, Didcot, OX11 0RQ, UK

[c]Information Management Services, Silver Spring, MD 20904, USA

[d]Radiation Safety Research Center, Nuclear Technology Research Laboratory, Central Research Institute of Electric Power Industry (CRIEPI), 2-11-1 Iwado-kita, Komae, Tokyo 201-8511, Japan

[1]All correspondence to Dr M P Little, Radiation Epidemiology Branch, Division of Cancer Epidemiology and Genetics, National Cancer Institute, NIH, DHHS, 9609 Medical Center Drive, Rockville, MD 20892-9778, USA, tel +1 301 875 3413 email: mark.little@nih.gov

**Short title:** Pneumonia after bacterial/viral infection and radiotherapy - implications for COVID-19 pneumonia

**Word count** (excluding references): 3611

**Abstract word count:** 300

2 tables, 7 figures, 15 Appendix Tables, 36 references

**Abstract**

Currently, there are about 14 ongoing clinical studies on low dose radiotherapy (LDRT) for COVID-19 pneumonia. One of the underlying assumptions is that irradiation of 0.5–1.5 Gy is effective at ameliorating viral pneumonia. Its rationale, however, relies on early human case series or animal studies mostly obtained in the pre-antibiotic era, where rigorous statistical analyses were not performed. It therefore remains unclear whether those early data support such assumptions. With standard statistical survival models, and based on a systematic literature review, we re-analyzed fourteen radiobiological animal datasets published in 1925-1973 in which animals received mostly fractionated doses of radiation before or after bacterial/viral inoculation, and assessing various health endpoints (mortality, pneumonia morbidity). In most datasets absorbed doses did not exceed 7 Gy. Various different model systems (rabbits, guinea pigs, dogs, cats, mice) and types of challenging infection (bacterial, viral) are considered. For seven studies that evaluated post-inoculation radiation exposure (more relevant to LDRT for COVID-19 pneumonia) the results are heterogeneous, with two studies showing a significant increase ($p<0.001$) and another showing a significant decrease ($p<0.001$) in mortality associated with radiation exposure. Among the remaining four studies, mortality risk was non-significantly increased ($p>0.3$) in two studies and non-significantly decreased in two others ($p>0.05$). For pre-inoculation exposure the results are also heterogeneous, with six (of eight) datasets showing a significant increase ($p<0.01$) in mortality risk associated with radiation exposure and the other two showing a significant decrease ($p<0.05$) in mortality (and in one case pneumonitis morbidity) risk. Collectively, these data do not provide clear support for reductions in morbidity or mortality associated with post-infection radiation exposure. For pre-infection radiation exposure the inconsistency of direction of effect makes this body of data difficult to interpret. Nevertheless, one must be cautious about adducing evidence from the published reports of these old animal datasets.

**Introduction**

Low dose radiotherapy (LDRT) for Coronavirus Disease 2019 (COVID-19) pneumonia was proposed in early April 2020 [1, 2]. At least 14 clinical studies are currently ongoing in 7 countries [3]. The rationale for clinical benefit, in other words the effectiveness of irradiation at the level of 0.5–1.5 Gy in treating viral pneumonia, largely relies on early human case studies or animal studies mostly obtained in the pre-antibiotic era, when a number of attempts were made to treat various non-cancer diseases with ionizing radiation, including virally- or bacterially-associated pneumonia. An influential paper underlying a number of proposals made for use of LDRT to treat COVID-19 pneumonia [1, 2] was Calabrese and Dhawan in 2013 [4] who reviewed 17 papers (all published before 1945) describing about 15 mostly relatively small case series, describing outcomes from low dose radiotherapy with X-rays (LDRT) for pneumonia. Their sampling framework and doses used are unknown, and therefore they are subject to ascertainment bias and are effectively uninterpretable. Calabrese and Dhawan [4] also identified four radiobiological animal studies of post-inoculation LDRT, all from experiments done in the 1940s, namely Fried [5] using a guinea pig model, Lieberman *et al.* [6] using a canine model, Baylin *et al* [7] using a cat model, and Dubin *et al.* [8] using a murine model, the first two of these for bacterially-induced pneumonia and the last two for virally-induced pneumonia. However, Calabrese and Dhawan [4] did not consider four other radiobiological studies relating to post-inoculation RT, nor results of eight others relating to pre-inoculation RT, and did not attempt any statistical reanalysis of these old data.

The aim of the present paper is to look at the totality of published radiobiological data relating to radiation exposure before or after inoculation with a viral or bacterial agent likely to result in pneumonia. Because of the age of the data being considered there are shortcomings in the original statistical analysis that was performed – indeed in all but a few cases [9, 10] there was no formal statistical analysis in the original reports. It is the purpose of this paper to report reanalysis of the data abstracted from the original publications so far as that is achievable, using standard statistical survival models in order to assess modification of pneumonia morbidity or mortality risk by radiation exposure before or after inoculation.

**Methods**

We aimed to capture all radiobiological datasets relating to moderate or LDRT whether given before or after viral or bacterial inoculation leading to pneumonia. We searched literature by means of a PubMed search (using terms ((radiation OR radiotherapy) AND pneumonia AND viral AND animal) OR ((radiation OR radiotherapy) AND pneumonia AND bacterial AND animal)) conducted on 2020-8-8, which returned 184 articles. We also searched for citations of the articles of Fried [5], Lieberman *et al.* [6], Baylin *et al* [7], Dubin *et al.* [8], and an authoritative contemporary review (of 1951) by Taliaferro and Taliaferro [11] on the same date. We did not restrict by date or language of the publication. We selected from these searches all relevant articles with information on radiobiological animal experiments in which there was any type of ionizing radiation exposure with determination of mortality or morbidity from bacterially- or virally-induced pneumonia. The datasets used are listed in Table 1. It should be noted that the datasets we used include three of the four cited by Calabrese and Dhawan [4], but did not include the paper of Fried [5] which we judged did not contain any quantitatively useful information. In Appendix A we provide details of the process used to abstract data from the publications that we identified as being potentially informative. The data was abstracted independently three times by MPL, WZ and RvD. We convert the given free-in-air dose in legacy units radiation absorbed dose (rad), roentgen (R) or rep in all studies to absorbed dose in Gray (Gy) via the scaling 1 R/rep =0.00877 Gy. 1 rad = 0.01 Gy [12].

*Statistical methods*

Details of the statistical models fitted are given in Table 1, and some further details on adjustments used are also given in the summary Table 2. Mortality and morbidity risks in the radiobiological cohorts of Murphy and Sturm [13], Lieberman *et al.* [6] and Dubin *et al.* [8] were assessed using a Cox proportional hazards models [14], with time after radiation exposure, if that followed the inoculation, or time after bacterial or viral inoculation, if that followed the radiation exposure, as timescale, in which the relative risk (RR) (=hazard ratio) of death for animal $i$ at time $a$ after inoculation was given by a linear model in dose:

$$RR_i[a, D_i | \alpha] = [1 + \alpha D_i] \tag{1}$$

or alternatively using a log-linear model in dose:

$$RR_i[a, D_i | \alpha] = \exp[\alpha D_i] \tag{2}$$

where $D_i$ is the total dose (in Gy), $\alpha$ is the excess relative risk coefficient (ERR) per unit dose (Gy). For the dataset of Murphy and Sturm [13] the only usable information available is the fact of irradiation of the mice (yes or no), so that the model fitted is of the form:

$$RR_i[a, 1_{exposure_i} | \alpha] = \exp[\alpha 1_{exposure_i = yes}] \tag{3}$$

For most of the other datasets a linear logistic model is fitted to the data (generally on number of animals that died in each group):

$$P_i[a, D_i | \alpha_0, \alpha_1] = \exp[\alpha_0][1 + \alpha_1 D_i] / [1 + \exp[\alpha_0][1 + \alpha_1 D_i]] \tag{4}$$

In some cases the more standard loglinear logistic model is fitted to the data (generally on number of animals that died in each group):

$$P_i[a, D_i | \alpha_0, \alpha_1] = \exp[\alpha_0 + \alpha_1 D_i]/[1 + \exp[\alpha_0 + \alpha_1 D_i]] \tag{5}$$

For the data of Fried [15], numbering only 7 animals and using as outcome improvement in pneumonia in relation to unirradiated controls, an exact logistic model was used [16], as non-exact methods did not converge. It is well known that the excess odds ratio (EOR) approximates to the excess relative risk [17]. All confidence intervals (CIs) and two-sided *p*-values are profile-partial-likelihood based [18]. In the murine dataset of Dubin *et al.* [8] in various subgroups risks were assessed in relation to radiation dose administered after inoculation or dose before inoculation. In the murine dataset of Quilligan *et al* [19] pre-inoculation dose was given to all animals. There is some uncertainty associated with the number of mice in the first of the control groups in this dataset, so a range is employed, spanning the plausible range of 6-10 mice, with 8 as the central estimate (Table 2). The model was stratified by the three experiments reported in the data of Dubin *et al.* [8] and by the three groups used by Lieberman *et al.* [6]. Tables B3 and B8 and Figure 1 and 2 show the risks in relation to dose for these two datasets. In various other datasets adjustment was made for various other covariates, as detailed in Tables B1-B15. In the fits to the pneumonia intensity data of Baylin *et al.* [7] we used either loglinear logistic regression (as described above) comparing each pneumonia intensity group and those with greater intensity vs every group with reduced intensity and because of the small number of animals (22) we also employed exact logistic methods [16]; we also used ordinal regression with log-linear link [20] fitting to all the ordered intensity groups. In fits of the days of infection data of Baylin *et al.* [7] we used a linear regression model, estimating the CIs via the bias-corrected advanced method [21]. All models were fitted via Epicure [22], R [23] or LogXact [16]. Two-sided levels of statistical significance are reported in all cases, with a conventional threshold for type I error of 2-sided $p<0.05$ used to assign statistical significance. All statistical analyses were independently performed by MPL and WZ to check for concordance. All datasets and analysis files are available in online supporting information (Appendix C).

*Author contributions*

MPL, WZ, RvD were responsible for data curation. MPL and WZ were responsible for statistical analysis. MPL and NH assembled the initial draft paper, and all authors contributed to the writing of the paper.

# Results

We present results of analyses of risk in relation to whether radiation exposure occurred after inoculation or before inoculation. The results are given in summary form in Table 2, which also provides summary details of the models used and assumptions made in fitting, and in more detail in Appendix B Tables B1-B15.

*Irradiation after inoculation*

Table 2 (and Table B1) demonstrates that there is a highly significant ($p<0.001$) increased risk of death associated with X-ray exposure after inoculation with *Pneumococcus* in the murine data of Murphy and Sturm [13], with RR = 3.67 (95% CI 1.84, 7.61). Table 2 (and Table B2) shows that there are weak indications ($0.05 < p < 0.1$) of decreased risk of pneumonia with post-inoculation dose in the dataset of Fried [15], whether for all guinea pigs or restricting to the six guinea pigs receiving *Staphylococcus aureus* inoculation. There is a highly significant decreasing trend ($p<0.001$) of mortality with post-inoculation dose in the dataset of Lieberman *et al.* [6] with EOR per Gy = -0.23 (95% CI -0.24, -0.16) (Table 2, Table B3), as also shown by Figure 1. However, Table B3 shows that this is largely driven by a single group, group 3, as also shown by Figure 2. The three groups in the study of Lieberman *et al* [6], were treated with slightly different X-ray energies, 80 kVp, 135 kVp and 200 kVp, respectively, and mean doses also slightly differed, 0.947 Gy, 1.639 Gy and 2.027 Gy, respectively (Table B3).

There are few indications of trend of degree of pneumonia infection with dose in the feline data of Baylin *et al.* [7], whether using logistic, exact logistic or ordinal models (Table 2, Table B4). However, Table B5 indicates that there is a significant decreasing trend of days of acute infection with dose in this dataset, with days of infection / Gy changing by -2.56 (95% CI -4.59, -0.33) ($p$=0.015), i.e., duration of infection decreasing with dose, as also shown by Figure 3.

There is a highly significant increasing trend ($p$<0.001) of mortality risk with dose after endemic *coccobacillus* infection in the rat data of Bond *et al.* [24], whether adjusting for likelihood of infection or not (Table 2, Table B6), with EOR per Gy = 0.81 (95% CI 0.72, 0.91), as also shown by Figure 4.

*Radiation administration before and after inoculation*

There is no significant trend ($p$>0.4) with post-inoculation dose in the data of Dubin *et al.* [8], whether using linear or log-linear models, which is confirmed also by Figure 1 (Table B7). However, there is a borderline significant decreasing trend ($p$=0.029) of mortality with pre-inoculation dose in this dataset, with EOR/Gy = -0.62 (95% CI -0.90, -0.09) (Table 2, Table B7) again confirmed by Figure 1.

*Irradiation before inoculation*

There is a non-significant positive trend with pre-inoculation dose ($p$>0.4) in the murine data of Tanner and McConchie [25] (Table 2, Table B8, Figure 5); results did not appreciably vary with the type of model used (linear logistic, log-linear logistic) or whether or not adjustment was made for the virus concentration (Table B8).

In the murine data of Beutler and Gezon [26] there are highly significant (all $p<0.005$) increasing trends of mortality with post-inoculation dose, whether in relation to mouse-adapted or egg-adapted PR8 influenza A virus and irrespective of the type of statistical model (linear logistic, loglinear logistic) used; for example with a linear logistic model the EOR per Gy is 0.16 (95% CI 0.06, 0.30) (Table 2, Table B9), and as also shown in Figure 6. The morbidity trends exhibit more heterogeneity, so that for the mouse-adapted virus the trends are generally negative, so that for example with a linear logistic model the EOR per Gy is -0.17 (95% CI -0.18, -0.15) (Table B9); however, for the egg-adapted virus the trends are generally positive with dose, so that for example with a linear logistic model the EOR per Gy is 0.41 (95% CI 0.14, 0.85) (Table B9), as also shown in Figure 6.

In the Swiss mice data of Hale and Stoner [27] there is no overall trend ($p>0.2$) of mortality with radiation dose given before inoculation with type III *pneumococcus*. However, if attention is restricted to the animals that received inoculation there is a highly significant increasing trend with dose ($p<0.001$), so that the EOR per Gy is 1.40 (95% CI 0.39, 5.47) (Table 2, Table B10). The same researchers went on to study a wider range of infective agents in the same strain of mice, and observed a generally highly significant ($p<0.005$) increase in mortality risk associated with radiation before inoculation with influenza virus, *pneumococcus* type III bacterial infection or *Trichinella spiralis* larval infection [28], whether adjusted or not for type of first immunizing infection, so that for example without such adjustment and excluding the *Trichinella spiralis* challenge infections the EOR/Gy = 1.40 (95% CI 0.39, 5.47) (Table B11, Table 2).

There is a highly significant ($p<0.001$) increase in mortality risk in the C57BL mouse data of Quilligan *et al* [19] associated with post-influenza-inoculation radiation dose with EOR/Gy ranging from 3.73 (95% CI 0.42, 85.85) to 4.75 (95% CI 0.56, 108.10) depending on how many mice are assumed to be in the first control group (Table B12, Table 2).

Pneumonitis morbidity and mortality were significantly decreased ($p<0.001$) after 3.5 Gy whole body air dose exposure in adult male albino CF-1 mice in the data of Berlin [9], so that for pneumonitis morbidity the EOR/Gy = -0.24 (95% CI -0.28, -0.17) and for pneumonitis mortality the EOR/Gy = -0.21 (95% CI -0.26, -0.14) (Table B13, Table 2). In contrast Table 2 (see also Table B14) and Figure 7 show reanalysis of slightly later data of Berlin and Cochran [10], which exhibits slightly heterogeneous results, with one set of experiments (given in Table III of Berlin and Cochran [10]) indicating a highly significant increase ($p=0.009$) in influenza mortality, whether or not adjusted for mode of administration of virus, but a different experimental set (reported in Table II of the paper) showing no significant effect ($p>0.1$) of radiation exposure on influenza morbidity or mortality. These experiments use a similar murine system, also given 3.5 Gy whole body air-dose exposure, as in the earlier paper of Berlin [9]. Lundgren *et al* [29] used a novel type of radiation exposure, aerosolized $^{144}$CeO$_2$, which delivers localized $\beta$ dose to the lungs of C57BL/6J mice. There was a small but highly significant increase in mortality risk associated with radiation exposure, with EOR/Gy = 0.008 (95% CI 0.002, 0.019, $p=0.002$) (Table 2, Table B15).

**Discussion**

We have re-analyzed fourteen radiobiological animal datasets, dating from the mid 1920s to the early 1970s, in which bacterial or viral agents were administered to induce pneumonia in animals that were also exposed to varying fractionated doses of radiation before or after inoculation. The statistical analysis in the original papers was limited, indeed in all but two cases [9, 10] there was no formal statistical analysis in the publications. We therefore judged it necessary to statistically reanalyze the data from the original publications with standard statistical models.

For the seven studies that evaluated post-inoculation radiation exposure (which is more relevant to LDRT for COVID-19 pneumonia) the results are heterogeneous, with two studies, those of Murphy and Sturm [13] and Bond *et al* [24], showing a significant increase ($p<0.001$) in mortality associated with radiation exposure, and another, that of Lieberman *et al* [6], showing a significant decrease ($p<0.001$) in mortality associated with radiation exposure. Among the remaining four studies, mortality risk was non-significantly increased in two studies, those of Baylin *et al* [7] and Tanner and McConchie [25] ($p=0.358$, $p=0.469$, respectively), and non-significantly decreased in two others, those of Fried [15] and Dubin *et al* [8] ($p=0.075$, $p=0.451$, respectively). Risks were only elevated in the third of the three groups studied by Lieberman *et al* [6]; the groups were treated with slightly different X-ray energies, and mean doses also slightly differed (see Results and Table B3). It is possible that these variations in mean dose and radiation energy may have some bearing on the differences observed (Table B3). For pre-inoculation exposure the results are also heterogeneous, with six (of eight) datasets showing significant increase in mortality risk associated with radiation exposure, namely those of Beutler and Gezon [26], Hale and Stoner [27], Hale and Stoner [28], Berlin and Cochran [10], Quilligan *et al* [19] and Lundgren *et al* [29] ($p<0.001$, $p<0.001$, $p<0.001$, $p=0.009$, $p<0.001$, $p=0.002$, respectively), and the other two, those of Dubin *et al* [8], and Berlin [9] showing a significant decrease in mortality (and in one case pneumonitis morbidity) risk ($p=0.029$, $p<0.001$, respectively). There was no clear systematic explanation for the heterogeneity in direction of effects, but reasons could include the different model systems (rabbits, dogs, cats, mice) being used, also possibly because of the various types of challenging infection, which included both bacterial agents (*Pneumococcus* types I, III,

*Staphylococcus aureus haemolyticus*, *Coccobacillus*) and viral ones (swine influenza, feline, Thylers mouse encephalitis, influenza type A, CAM A-prime influenza, influenza A/PR8, and influenza $A_0$) (see Table 1). It is possible that the range of doses used, and the variable degree of fractionation employed may also be factors, although there does not appear to be an obvious pattern linking these to the direction or strength of effect, as shown by Tables 1 and 2. More recently, Hasegawa *et al* [30] showed that irradiation before influenza vaccination and a succeeding lethal challenge influenza inoculation exacerbates mortality from influenza, but that vaccination prior to irradiation confers protection against a subsequent challenge influenza inoculation. Dadachova *et al* [31] reported that targeted radionuclide immunotherapy induces *Streptococcus pneumoniae* killing *in vivo*, thereby alleviating bacterial pneumonia. As can be inferred, both these publications [30, 31] address somewhat different questions to those of this paper.

The major strength of the present analysis is that we use standard statistical models to assess the totality of published radiobiological data on LDRT given either before or after virally- or bacterially-induced pneumonia. Moreover, unlike the review of Calabrese and Dhawan [4] we have undertaken a systematic review of the literature. The data we used was independently abstracted from the published reports by three of the authors. A significant weakness is the heterogeneity in the study designs, both the animal model systems and the infective agents used, alluded to above. There is some uncertainty as to precisely what experimental procedures were followed in some of these old datasets and one cannot be sure that the experimenters in these studies were blinded to the exposure status of the animals. This is exemplified by the study of Lieberman *et al* [6], where in the third group, most recovered animals received irradiation at 3 days after bacterial inoculation, which exceeded the average lifetime of 2.1 days in control (infected but not irradiated) animals: this would suggest an experimental selection bias. It is also not always clear what the disease endpoints were, as for example in the data of Bond *et al* [24]. Last, about a third (5/14) of the radiobiological datasets we analyzed here dealt exclusively with bacterial pneumonia, which is less relevant to discussion of LDRT for COVID-19 pneumonia, due to a significant difference in the pathogenesis of bacterial and viral pneumonia. Reliant as we were on electronic publication databases (in particular PubMed, ISI Thompson) it is possible that our literature search could have missed some relevant older datasets, given the incompleteness in coverage of publications 80 or more years previously.

Altogether, early radiobiology data do not suggest that there are strong variations in mortality or morbidity following radiation exposure after bacterial or viral inoculation. In particular, the heterogeneity in the results of our statistical analysis suggest that these early datasets do not serve as strong supportive evidence that LDRT of infected individuals reduces mortality. Although there are stronger indications of modifications of risk by radiation exposure before inoculation, the inconsistency of direction of effect makes this body of data difficult to interpret and has little relevance to LDRT for COVID-19 pneumonia.

Rödel *et al* [32] reviewed some of the human epidemiological and radiobiological literature on LDRT. While acknowledging limitations in understanding the possible mechanism, Rödel *et al* [32] suggested that LDRT may stimulate anti-viral immunity via the modulating effects of type I interferons in the early stages of SARS-CoV-2 infection. Rödel *et al* [32] concluded that LDRT with a single dose of 0.5 Gy to the lungs warranted clinical investigation, while acknowledging the need for strict monitoring and disease phase-adapted treatment based on lung function tests and clinical markers (e.g. IL-6 and D-dimer in serum). Schaue and McBride [33] echoed some of the concerns of Rödel *et al* [32] on the importance of correctly timing the use of LDRT in treatment of SARS-CoV-2, but were much more cautious, and suggested that, for example, it was unlikely that LDRT would effectively counter the virally-induced cytokine storm that is a feature of the more severe forms of infection. Schaue and McBride [33] and even more forcefully Kirsch *et al* [34] suggested that the known deleterious adverse late health effects of 0.5-1.5 Gy administered to the lungs via increased risk of cancer and circulatory disease [35, 36] must be weighed against the uncertain therapeutic benefits of LDRT. Kirsch *et al* concluded that "based on the available data, the potential risks of such LDRT trials outweigh the potential benefits" and recommended that "further preclinical work is needed to demonstrate efficacy of radiotherapy to provide scientific justification for a clinical trial in patients with COVID-19" [34].

Collectively, these animal data do not provide clear support for reductions in morbidity or mortality associated with post-infection radiation exposure. For pre-infection radiation exposure, the inconsistency of direction of effect makes this body of data difficult to interpret. Nevertheless, one must be cautious about adducing evidence from the published reports of these old animal datasets.

## Acknowledgements

This work was funded by the Intramural Research Program of the Division of Cancer Epidemiology and Genetics, National Cancer Institute, National Institutes of Health.15

Table 1. Radiobiological animal data used for re-analysis of effects of low dose radiotherapy on bacterially- or virally-induced pneumonia

| Author | Ref. | Animal strain | Infective agent(s) | Endpoint(s) analyzed | Mean (range) cumulative dose (Gy) | Number of fractions | No of animals | Statistical model used for re-analysis |
|---|---|---|---|---|---|---|---|---|
| *Radiation exposure after inoculation* | | | | | | | | |
| Murphy and Sturm | [13] | Rabbits | Type I *Pneumococcus* | Mortality | NA | 5 / day | 75 | Loglinear Cox |
| Fried | [15] | Guinea pig | *Staphylococcus aureus haemolyticus* | Improvement in pneumonia morbidity | 0.357 (0-0.833) | 1 | 7 | Exact logistic |
| Lieberman *et al.* | [6] | Dogs | Type I+III *Pneumococcus* | Mortality | 1.513 (0.0-4.096) | 1-3 | 45 | Linear + loglinear Cox |
| Baylin *et al.* | [7] | Cats | Feline virus (Baker) | Degrees of pneumonia | 0.598 (0.0-1.754) | 1-2 | 22 | Loglinear logistic, exact logistic + ordinal |
| Tanner and McConchie | [25] | CFW mice | Theilers FA mouse encephalitis virus | Mortality | 3.114 (0.0-5.262) | 6 | 196 | Linear binomial logistic |
| Bond *et al.* | [24] | Sprague Dawley rats | Endemic *Cocobacillus* | Mortality | 3.912 (0.0-8.770) | 1 | 1059 | Loglinear binomial logistic |
| *Radiation exposure before and after inoculation* | | | | | | | | |
| Dubin *et al.* | [8] | White mice | Swine influenza virus | Mortality | 0.366 (0.0-1.754) | 1-3 | 252 | Linear + loglinear Cox |
| *Radiation exposure before inoculation* | | | | | | | | |
| Beutler and Gezon | [26] | Germantown white mice | PR8 strain type A influenza virus (mouse adapted, egg-adapted) | Mortality and morbidity | 1.682 (0-5.262) | 1 | 1635 | Linear + loglinear binomial logistic |
| Hale and Stoner | [27] | Swiss mice | Type III *Pneumococcus* | Mortality | 2.443 (0-5.701) | 1 | 140 | Linear + loglinear binomial logistic |
| Hale and Stoner | [28] | Swiss mice | Influenza type A, *Trichinella spiralis*, type III *Pneumococcus* | Mortality | 3.112 (0.0-6.139) | 1 | 658 | Linear binomial logistic |
| Quilligan *et al.* | [19] | C57BL male mice | PR8 strain type A influenza virus | Mortality | 6.331 (0.0–14.471) | 15 | 30 - 34 | Linear binomial logistic |
| Berlin | [9] | CF-1 adult albino male mice | CAM A-prime strain influenza virus | Morbidity+mortality | 1.594 (0.0-3.070) | 1 | 362 | Linear binomial logistic |
| Berlin and Cochran | [10] | CF-1 adult albino male mice | PR8 strain type A influenza virus | Morbidity+mortality | 1.797 (0.0-4.385) | 1 | 660 | Linear binomial logistic |
| Lundgren *et al.* | [29] | Female C57BL/6J mice | Type $A_0$ influenza virus | Mortality | 88.763 (0.0-190.0) | Continuous | 364 | Linear binomial logistic |

NA, not available.

**Table 2. Summary of modifying effects of pre-inoculation or post-inoculation radiation exposure on bacterially- or virally-induced pneumonia in re-analyzed radiobiological data**

| Author | Ref. | Infective agent(s) | Endpoint(s) analyzed | Mean (range) cumulative dose (Gy) | Excess relative risk Statistical model used for re-analysis | Other notes on regression | Excess relative risk / Gy, excess odds ratio / Gy (+95% CI) | $p$-value |
|---|---|---|---|---|---|---|---|---|
| | | | | Radiation exposure after inoculation | | | | |
| Murphy and Sturm | [13] | Type I *Pneumococcus* | Mortality | NA | Loglinear Cox | No dose – fact of irradiation | 3.67 (1.84, 7.61)[a] | <0.001 |
| Fried | [15] | *Staphylococcus aureus haemolyticus* | Improvement in pneumonia morbidity | 0.357 (0-0.833) | Exact logistic | Only animals receiving inoculation | 2.42[b] (-0.46[c], +∞[c]) | 0.075 |
| Lieberman et al. | [6] | Type I+III *Pneumococcus* | Mortality | 1.513 (0.0-4.096) | Linear Cox | Model stratified by three groups | -0.23 (-0.24, -0.16) | <0.001 |
| Baylin et al. | [7] | Feline virus (Baker) | Degrees of pneumonia | 0.598 (0.0-1.754) | Ordinal logistic | | 0.55 (-0.62, 1.76) | 0.358 |
| Tanner and McConchie | [25] | Theilers FA mouse encephalitis virus | Mortality | 3.114 (0.0-5.262) | Linear binomial logistic | | 0.09 (-0.09, 0.61) | 0.469 |
| Bond et al. | [24] | Endemic *Cocobacillus* | Mortality | 3.912 (0.0-8.770) | Loglinear logistic | | 0.81 (0.72, 0.91) | <0.001 |
| Dubin et al. | [8] | Swine influenza virus | Mortality | 0.366 (0.0-1.754) | Linear Cox | Model stratified by three experiments | -0.13 (-0.35, 0.27) | 0.451 |
| | | | | Radiation exposure before inoculation | | | | |
| Dubin et al. | [8] | Swine influenza virus | Mortality | 0.366 (0.0-1.754) | Linear Cox | Model stratified by three experiments | -0.62 (-0.90, -0.09) | 0.029 |
| Beutler and Gezon | [26] | PR8 strain type A influenza virus (mouse adapted) | Mortality | 1.682 (0-5.262) | Linear binomial logistic | | 0.16 (0.06, 0.30) | <0.001 |
| Hale and Stoner | [27] | Type III *Pneumococcus* | Mortality | 2.443 (0-5.701) | Linear binomial logistic | Only animals with type III *Pneumococcus* administered | 1.40 (0.39, 5.47) | <0.001 |
| Hale and Stoner | [28] | Influenza type A, *Trichinella spiralis*, type III *Pneumococcus* | Mortality | 3.112 (0.0-6.139) | Linear binomial logistic | Adjusted for challenge infection type, fitted to experiments with influenza and *Pneumococcus* challenge infections only | 1.71 (0.97, 3.02) | <0.001 |
| Quilligan et al. | [19] | PR8 strain type A influenza virus | Mortality | 6.331 (0.0–14.471) | Linear binomial logistic | 8 mice in 1st control group assumed | 4.24 (0.49, 96.97) | <0.001 |
| Berlin | [9] | CAM A-prime strain influenza virus | Pneumonitis morbidity | 1.594 (0.0-3.070) | Linear binomial logistic | | -0.24 (-0.28, -0.17) | <0.001 |
| Berlin and Cochran | [10] | PR8 strain type A influenza virus | Mortality | 1.797 (0.0-4.385) | Linear binomial logistic | Analysis of influenza mortality adjusting for mode of administration of virus | 0.25 (0.05, 0.57) | 0.009 |
| Lundgren et al. | [29] | Type $A_0$ influenza virus | Mortality | 88.763 (0.0-190.0) | Linear logistic | binomial | 0.007 (0.002, 0.017) | 0.002 |

[a]relative risk among exposed animals
[b]median unbiased estimator
[c]exact 95% CI

**Figure 1. Dose response for mortality (+95% CI) in the murine data of Dubin *et al.* [8] and in the canine data of Lieberman *et al.* [6]**[1]

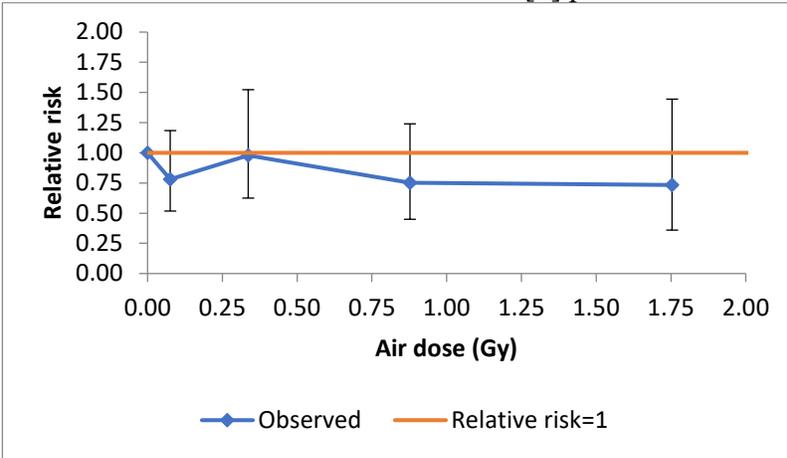

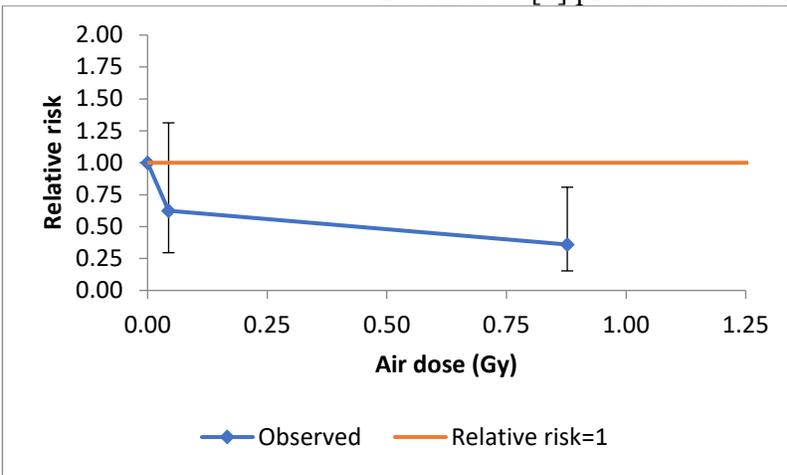

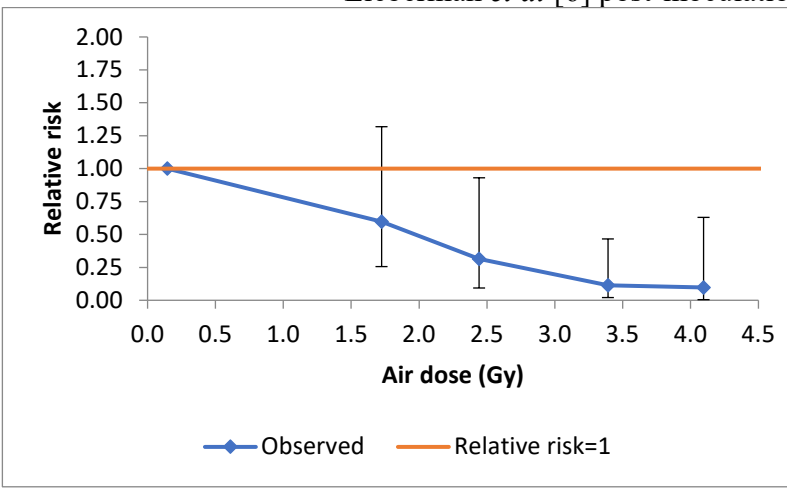

---

[1] Dubin *et al* breakpoints at 0.01, 0.2, 0.5, 1 Gy post inoculation dose, 0.01, 0.1 Gy pre-inoculation dose, Lieberman *et al* breakpoints at 1, 2, 3, 4 Gy

**Figure 2. Dose response for mortality (+95% CI) in the canine data of Lieberman *et al.* [6], by study group.**[2]

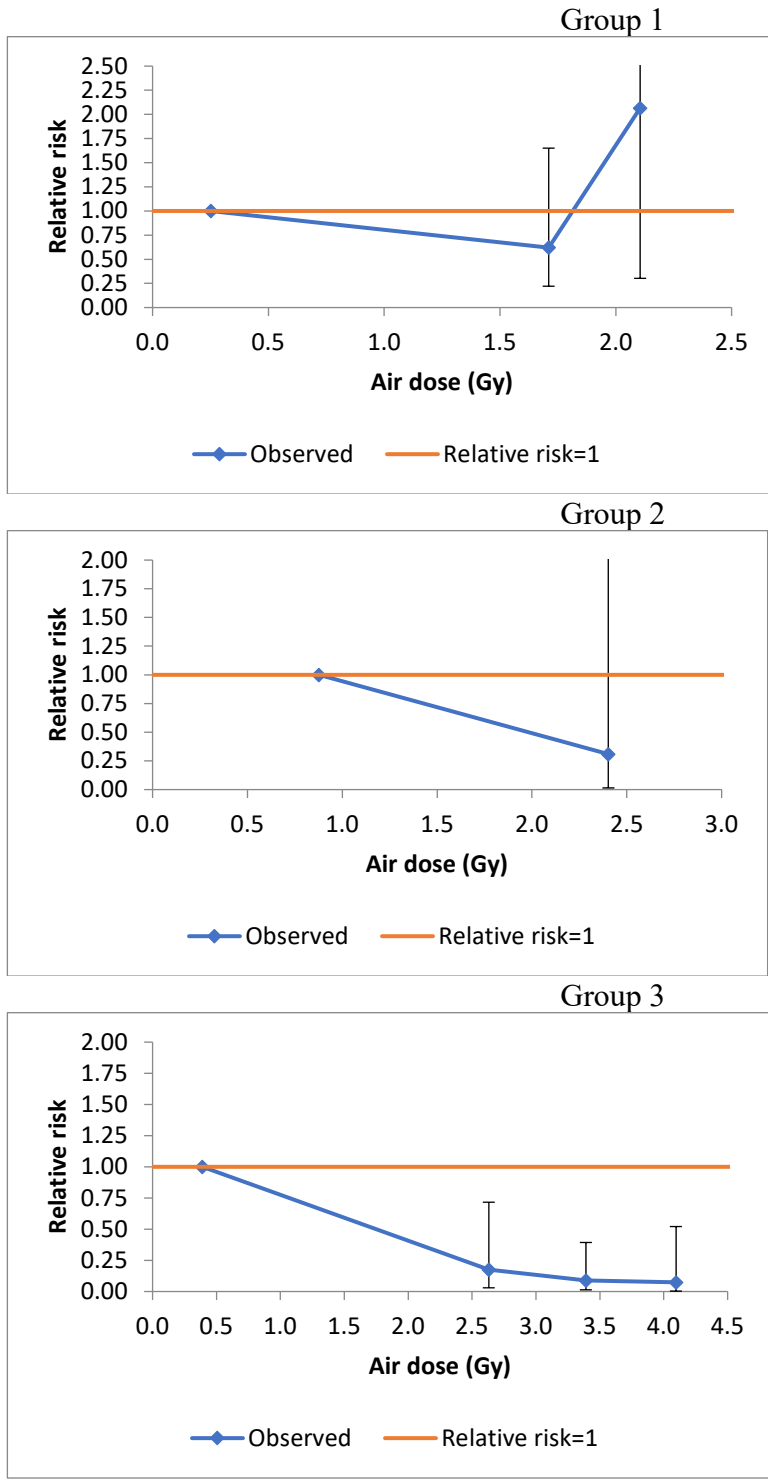

---

[2]Breakpoints at 1, 2 Gy [group 1], 2 Gy [group 2], 2, 3, 4 Gy [group 3]

**Figure 3. Days of acute pneumonia infection (+95% CI) vs post-inoculation dose in feline data of Baylin *et al* [7]**[3]

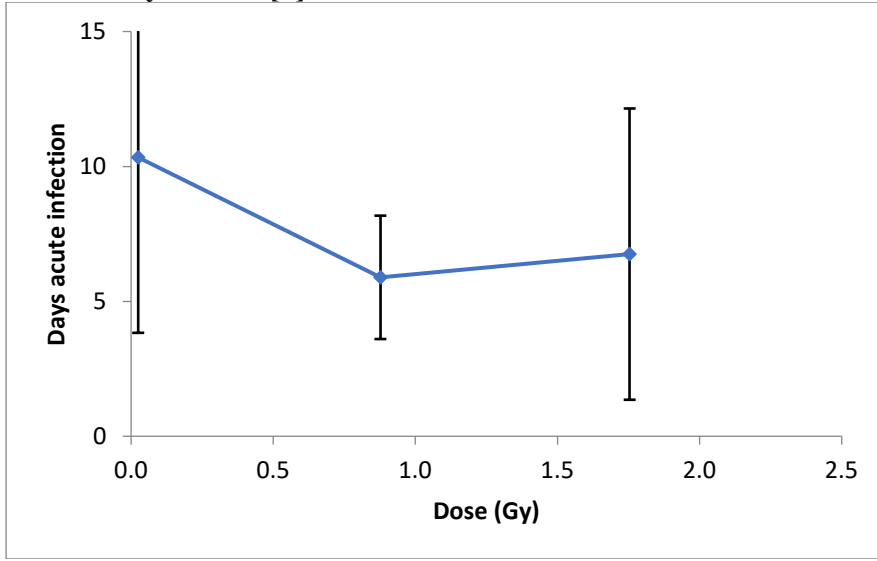

---
[3] Breakpoints at 0.75, 1.5 Gy

**Figure 4. Mortality in Sprague Dawley rats (+95% CI) after endemic *coccobacillus* infection and post-inoculation radiation exposure in data of Bond *et al* [24][4]**

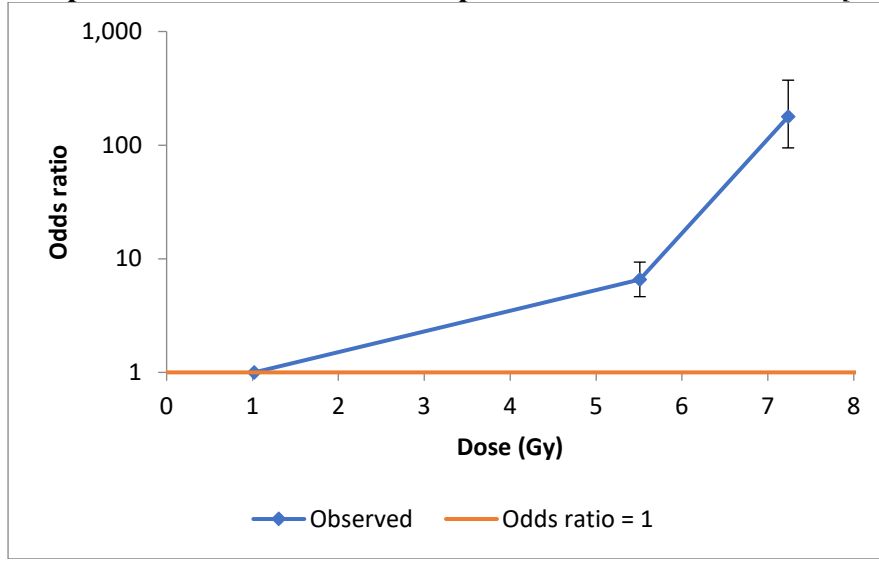

---
[4] Breakpoints at 5, 7 Gy

**Figure 5. Mortality in mice (+95% CI) after mouse encephalitis virus and post-inoculation radiation exposure in data of Tanner and McConchie [25][5]**

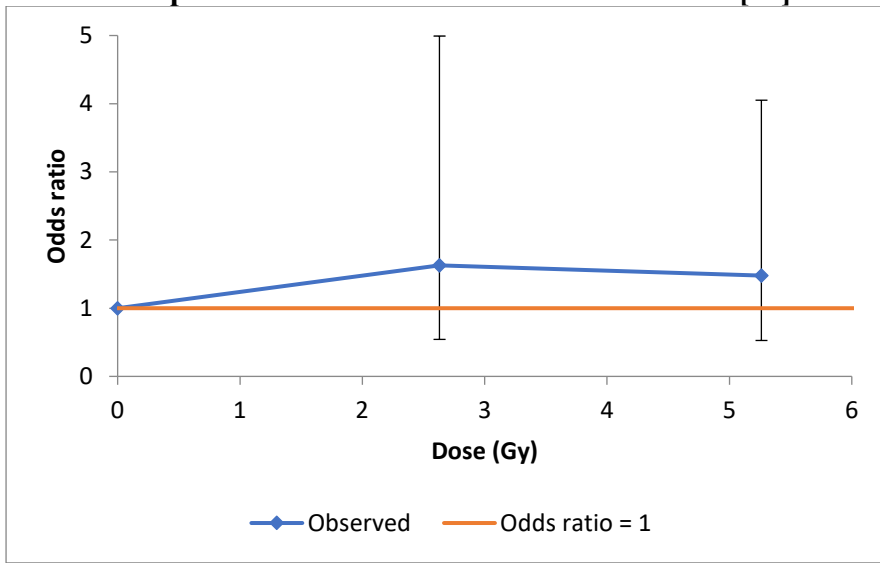

---
[5] Breakpoints at 2.5, 4.5 Gy

**Figure 6. Mortality and morbidity risks (+95% CI) in Germantown white mice associated with pre-inoculation radiation exposure to mouse-adapted or egg-adapted PR8 influenza A virus in the data of Beutler and Gezon [26]**[6]

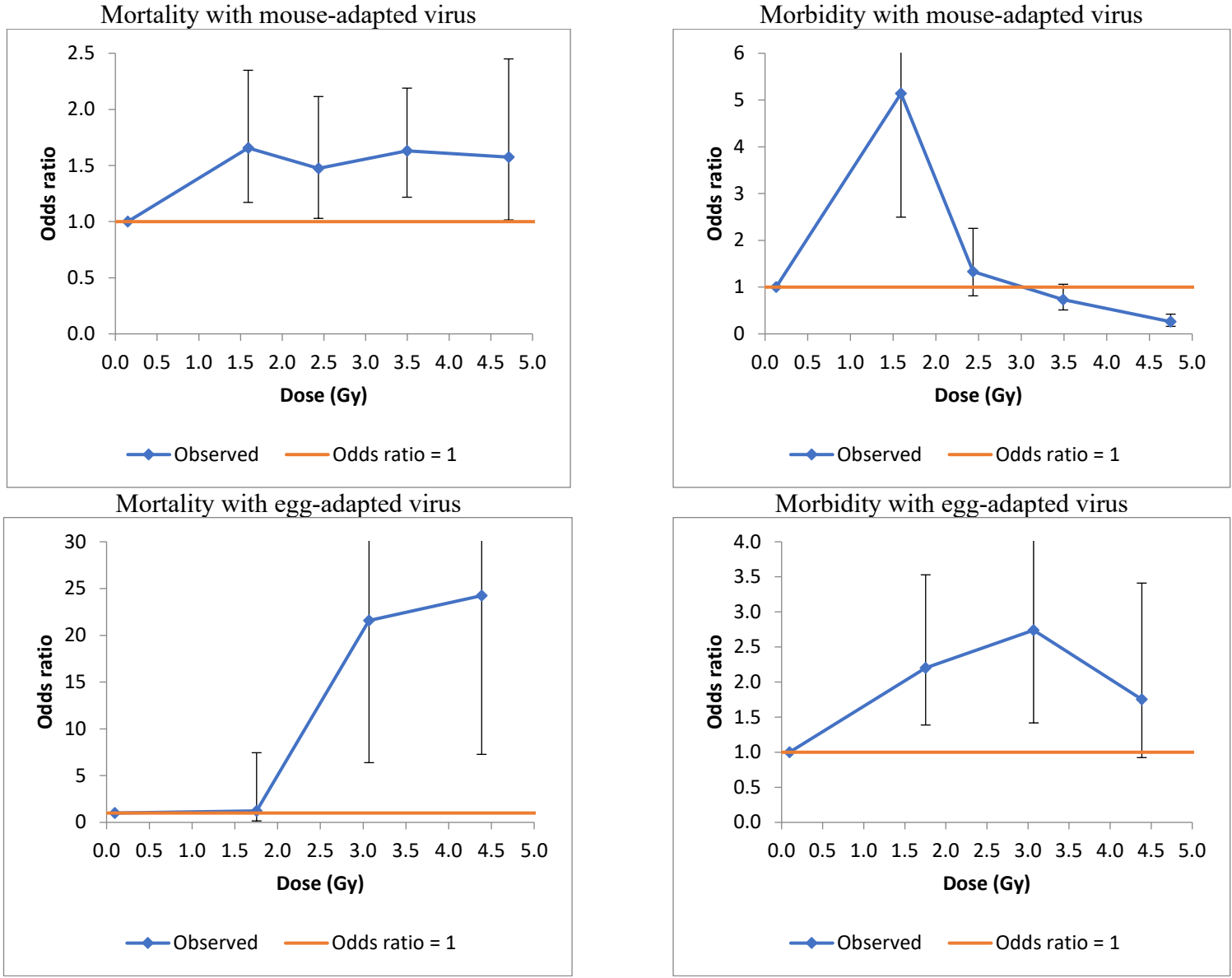

---

[6] Breakpoints at 1, 2, 3, 4 Gy [mouse-adapted virus, mortality+morbidity], 1.5, 3, 4 Gy [egg-adapted virus, mortality+morbidity]

**Figure 7. Influenza morbidity and mortality in mice (+95% CI) after PR8 strain type A influenza virus and pre-inoculation radiation exposure in data of Berlin and Cochran [10]**[7]

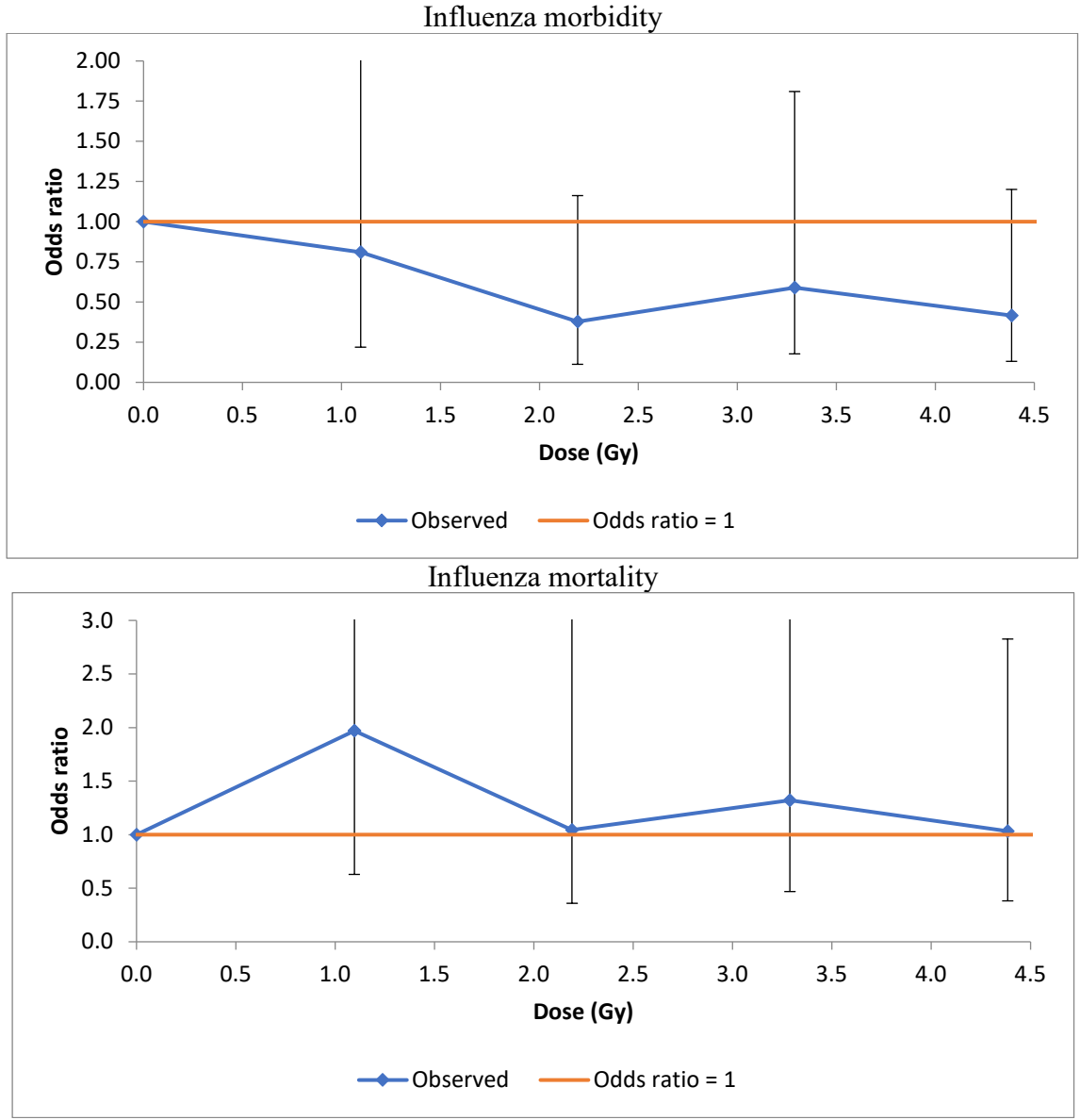

---

[7]Breakpoints at 1, 2, 3, 4 Gy

# Appendix A. Description of data extraction

In this appendix we outline the steps taken to transcribe data from the 14 papers relating to radiobiological animal experiments produced by our literature search.

## *Experiments with irradiation after inoculation*

**Murphy and Sturm [13]**

Data was transcribed from all three columns of text Figure 2 (p.252) into a table, giving information on group number, animal number, X-ray exposure status, heat exposure status, amount of serum, amount of culture, days survival. The days survival entry requires somewhat approximately translating the length of the given horizontal bars into survival length. This item was rounded to whole numbers (integers). There was information on 75 individual rabbits in this data.

**Fried [15]**

Data was transcribed from Table II (p.437) for experiment no 7 into a table, giving information on animal number, injection status, Radiation dose (Gy – converting 1 r = 0.00877 Gy), radiation time post injection, killed hours after injection, improvement (vs unirradiated controls) (coded as + or -), no improvement (vs unirradiated controls) (coded as + or -), doubtful improvement (vs unirradiated controls) (coded as + or -) and worsening (vs unirradiated controls) (coded as + or -).

**Lieberman *et al* [6]**

Data was transcribed from Tables 1-3 (pp.97-98) into a table giving dog number, group number, blood culture (coding "…." = -1, "+"=1, "++"=2, "+++"=3, "++++"=4, "0"=0) pneumonia type ("I"=1,"III"=3), radiation energy (keV)(from table title), radiation 1st day (Gy converting 1 r = 0.00877 Gy), radiation 2nd day (Gy converting 1 r = 0.00877 Gy), radiation 3rd day (Gy converting 1 r = 0.00877 Gy), radiation total (Gy converting 1 r = 0.00877 Gy), radiation total excluding day of death (Gy converting 1 r = 0.00877 Gy), death day, dead (coding 1=dead, 0=no death). For those animals that are indicated as having recovered (animal 519 in Table 2, animals 316, 383, 385, 481, 588 in Table 3) a large value (9999) was used as the end of follow-up time (Death day), with death coded as 0 for these 6 animals, so that the records for these animals only contribute to the denominator in the Cox model fit.

**Baylin *et al* [7]**

We constructed a table from Table 1 (p.474) with group number, cat number, area treated, degree of pneumonia, 24 h dose (Gy, calculated as 1 r = 0.00877 Gy) post-symptoms, 48 h dose (Gy, calculated as 1 r = 0.00877 Gy), 72 h dose (Gy, calculated as 1 r = 0.00877 Gy) post-symptoms, length of days in acute phase. It should be noted that in cats treated 24 h after onset of symptoms (Group II), when 100 r is given twice (cats 14, 16) that means 100 r (0.877 Gy) is given at 24 h and again at 48 h, so there were two separate entries in these columns. Likewise for cats treated 48 h after onset of symptoms (Group III) when 100 r (0.877 Gy) is given twice (cats 21, 22) then 100 r (0.877 Gy) is given at 48 h and again at 72 h after onset of symptoms.

**Tanner and McConchie [25]**

We constructed a table from Tables II and III (p.102) with numbers of LD-50 doses, radiation dose (Gy, calculated as 1 r = 0.00877 Gy), radiation portal, virus dilution, radiation start day, radiation end day, radiation spacing (days between fractions), number of mice, number died. It should be noted that the mice are irradiated daily, so for example in the 1st two groups mice receive 50 r (0.439 Gy) or 100 r (0.877 Gy) daily on days 1 through 6 after inoculation, so that cumulative doses of 300 r (2.631 Gy) or 600 r (5.262 Gy) are given to these groups and similarly for all other entries in Table II and Table III.

**Bond *et al* [24]**

We used a combination of the data in Table 1 (p.28) and Table 2 (p.29) to create a table with fields for experiment number, series, sex, age start weeks (minimum), age start weeks (maximum), infection status [coded as likely unexposed to infection, probably exposed to infection, highly probably exposed to infection – so that Experiments 1 and 2 were both coded as probably exposed to infection, and all except the last group of Experiment 3 were coded as highly probably exposed to infection, the last group of Experiment 3 being coded as likely unexposed to infection], minimum radiation dose (in Gy, using 1 r =0.00877 Gy), maximum radiation dose (in Gy, using 1 r =0.00877 Gy), deaths infected, deaths total, survived infected, survived total. These numbers were used to compute entries for numbers of deaths among infected and uninfected animals, which formed the basis of the analysis. The radiation dose used was the mean of the minimum and maximum doses.

*Experiments with irradiation before and after inoculation*

**Dubin *et al* [8]**

Data was transcribed on the experiment number, group number, mice per group, mouse number, dose (Gy, calculated as 1 r = 0.00877 Gy) 24 h post-inoculation, dose (Gy, calculated as 1 r = 0.00877 Gy) 48 h post-inoculation, dose (Gy, calculated as 1 r = 0.00877 Gy) 72 h post-inoculation, dose (Gy, calculated as 1 r = 0.00877 Gy), 48 h pre-inoculation dose (Gy, calculated as 1 r = 0.00877 Gy), start follow-up day (identically = 0), end follow-up day, death at end of follow-up (0=no, 1=yes). This is reconstructed from Table 1 (p.479). For example the first row of Table 1 suggests that there are 25 mice untreated (0 Gy), with cumulative mouse deaths of 2, 9, 13, 21, 22, 22, 22, 22 at 3, 4, 5, 6, 7, 8, 9, 10 days, implying there are 3 mice surviving to at least day 11 without dying, two mice dying at day 3, 7 mice dying on day 4, 4 mice dying on day 5, 8 mice on day 6, and a single mouse on day 7. We accordingly reconstructed individual animal records for all 25 mice in this group, as shown in this Table. We continued in this way for all elements of this Table. All information about doses at various times before or after inoculation came from the text description above each row in the Table. There was information on 252 individual mice in this data.

## *Experiments with irradiation before inoculation*

### **Beutler and Gezon [26]**

From Table II (p.232) we constructed a Table for mortality from mouse-adapted virus with records containing for each group of animals the experiment number, the suspension group, the negative log viral dilution, the group number (an arbitrary number, which we took to start at 1 for the first group, increasing by 1 for each successive group), the radiation dose (Gy, using 1 r = 0.00877 Gy), the number of mice that died and the number exposed.

Likewise we constructed a similar table for morbidity from mouse-adapted virus, based on Table III (p.233) with records containing for each group of animals the experiment number, the suspension group, the negative log viral dilution, the group number (an arbitrary number, which we took to start at 1 for the first group, increasing by 1 for each successive group), the radiation dose (Gy, using 1 r = 0.00877 Gy), the number of mice with morbidity and the number in each treatment group.

From Table V (p.237) we constructed a Table for mortality from egg-adapted virus with records containing for each group of animals the experiment number, the negative log viral dilution, the group number (an arbitrary number, which we took to start at 1 for the first group, increasing by 1 for each successive group), the radiation dose (Gy, using 1 r = 0.00877 Gy), the number of mice that died and the number in each treatment group.

Likewise from Table VI (p.237) we constructed a Table for morbidity from egg-adapted virus with records containing for each group of animals the experiment number, the negative log viral dilution, the group number (an arbitrary number, which we took to start at 1 for the first group, increasing by 1 for each successive group), the radiation dose (Gy, using 1 r = 0.00877 Gy), the number of mice with morbidity and the number in each treatment group.

**Hale and Stoner [27]**

From Table 1 (p.327) we produced a table with group number, number of mice, passively immunized (yes, no), radiation dose (Gy, using 1 rep =0.00877 Gy), pneumococcus injected intra-abdominally (yes, no), and number of deaths.

**Hale and Stoner [28]**

From Table I, II, III, V we produced a table with experiment (table), group, no of mice, 1st immunizing infection type, radiation dosage (Gy, calculated as 1 rep = 0.00877 Gy), type of 2nd challenge infection, days from first injection to second, deaths.

**Quilligan *et al* [19]**

We constructed a table from the text describing two series of experiments on p.509. "Another experiment with small numbers was carried out in which the animals were exposed to the 110 r/day dose for 15 days. At this time, these animals, together with a control group, were given live PR8 virus by the intraperitoneal route. Twelve days later, both groups of animals were challenged with the same virus by the intranasal route. Seven of the eight irradiated animals died, whereas there were no deaths in the nonirradiated vaccinated animals. A second experiment carried out in the same way showed 4/6 deaths in the vaccinated irradiated group and 1/10 deaths in the vaccinated nonirradiated group." The numbers in the first control group are not given, but presumably about 8 animals were used, our central estimate. As sensitivity analysis we also assumed there to be either 6 or 10 animals in this first control group. Roentgen (r) are converted to Gy via 1 r = 0.00877 Gy.

**Berlin [9]**

A table was constructed from Table I (p.865). The first table contained details on experiment, radiation dose (Gy, calculated as 1 R = 0.00877 Gy), number with pneumonitis morbidity, morbidity number at risk, number with pneumonitis mortality, and mortality number at risk.

**Berlin and Cochran [10]**

Separate tables were constructed from Table II (p.346) and Table III (p.347). The first table contained data on dilution of virus inoculum, dose of X ray (Gy, calculated as 1 R = 0.00877 Gy), number of mice, X-ray deaths, influenza morbidity, influenza mortality, mean survival time days.

The second table contained data on experiment number, route of viral exposure (intra-nasal, air), days between X-ray and inoculation, number of sham exposure influenza deaths, number of sham exposed mice, X-irradiated influenza deaths, number of X-irradiated mice, sham exposed days to death, sham exposed days to death standard error (SE), X-ray exposed days to death, and X-ray exposed days to death SE.

**Lundgren *et al* [29]**

We used Table I (p.240) to construct a table with weeks post $^{144}$CeO$_2$ inhalation, absorbed dose (Gy, calculated as 1 rad = 0.01 Gy), SD dose (Gy, calculated as 1 rad = 0.01 Gy), mortality in irradiated mice, number of irradiated mice, mortality in control mice, and numbers of control mice.

**Appendix B. More detailed Tables of results**
**Table B1. Relative risk of mortality in rabbits given X-rays after inoculation with type I *Pneumococcus* of Murphy and Sturm [13] via fit of a Cox proportional hazards model**

|  | Relative risk (hazard ratio) (+ 95% CI) | *p*-value |
|---|---|---|
| X-ray vs not | 3.67 (1.84, 7.61) | <0.001 |

**Table B2. Improvement in pneumonia morbidity in guinea pigs associated with radiation exposure post inoculation to *Staphylococcus aureus haemolyticus* in the data of Fried [15], via fit of a logistic model via exact methods**

|  | ln[EOR] / Gy + 95% CI | *p*-value |
|---|---|---|
| Only guinea pigs receiving inoculation | 2.42[a] (-0.46[b], +∞[b]) | 0.075 |
| All guinea pigs | 2.73[a] (-0.08[b], +∞[b]) | 0.057 |

[a]median unbiased estimator
[b]exact 95% CI

**Table B3. Risks of death associated with radiation exposure post-inoculation to types I+III *Pneumococcus* in the canine data of Lieberman *et al.* [6] by study group, via fit of a Cox proportional hazards model**

| Model | Group | Mean X-ray energy (kVp) | Mean dose (Gy) | ERR / Gy (+95% CI) | *p*-value |
|---|---|---|---|---|---|
| Linear | Group 1 | 80 | 0.947 | -0.05 (-0.35, 0.86) | 0.860 |
| Loglinear | Group 1 | | | -0.05 (-0.58, 0.50) | 0.861 |
| Linear | Group 2 | 135 | 1.639 | -0.37 (-0.41, 0.82) | 0.157 |
| Loglinear | Group 2 | | | -1.75 (-6.25, 0.27) | 0.096 |
| Linear | Group 3 | 200 | 2.027 | -0.23 (-0.24, -0.16) | <0.001 |
| Loglinear | Group 3 | | | -0.68 (-1.11, -0.29) | <0.001 |
| Linear | All groups | 80-200 | 1.513 | -0.23 (-0.24, -0.16) | <0.001 |
| Loglinear | All groups | | | -0.49 (-0.81, -0.19) | <0.001 |

**Table B4. Logistic regression of intensity of pneumonia infection following feline virus (Baker) inoculation in relation to post-inoculation dose in feline data of Baylin *et al* [7]**

| Group [pneumonia intensity group contrasts] | ln[OR] / Gy + 95% CI | *p*-value |
|---|---|---|
| *Loglinear logistic regression* | | |
| Very slight and above vs none | 21.29 (<-100, >100) | 0.173 |
| Slight and above vs everything below | 1.35 (-0.77, 4.77) | 0.235 |
| Slight to moderate and above vs everything below | 0.92 (-0.42, 2.47) | 0.181 |
| Moderate and above vs everything below | 0.34 (-1.03, 1.75) | 0.622 |
| Moderate to marked and above to everything below | -1.35 (-4.77, 0.77) | 0.235 |
| Marked vs everything below | 0.49 (-3.11, 4.12) | 0.755 |
| *Exact loglinear logistic regression* | | |
| Very slight and above vs none | 0.33[a] (-2.41[b], +∞[b]) | 0.818 |
| Slight and above vs everything below | 1.29[c] (-1.20[b], 5.73[b]) | 0.530 |
| Slight to moderate and above vs everything below | 0.88[c] (-0.61[b], 2.61[b]) | 0.314 |
| Moderate and above vs everything below | 0.32[c] (-1.20[b], 1.90[b]) | 0.846 |
| Moderate to marked and above to everything below | -1.29[c] (-5.73[b], 1.20[b]) | 0.530 |
| Marked vs everything below | 0.46[c] (-4.19[b], 5.12[b]) | 1.000 |
| *Ordinal regression* | | |
| | 0.55 (-0.62, 1.76) | 0.358 |

[a] median unbiased estimator
[b] exact 95% CI
[c] conditional maximum likelihood estimator

**Table B5. Linear regression of days of acute pneumonia infection following feline virus (Baker) inoculation vs post-inoculation radiation dose in data of Baylin *et al* [7]**

| days / Gy + 95% bootstrap CI | BCA Bootstrap *p*-value |
|---|---|
| -2.56 (-4.59, -0.33) | 0.015 |

**Table B6. Loglinear logistic regression of deaths from acute lung infection following endemic coccobacillus infection vs post-inoculation radiation dose in Sprague Dawley rat data of Bond *et al* [24]**

| Adjustment | EOR / Gy + 95% CI | *p*-value |
|---|---|---|
| Unadjusted | 0.81 (0.72, 0.91) | <0.001 |
| Adjusted for likelihood of infection | 0.85 (0.75, 0.95) | <0.001 |

**Table B7. Risks of death associated with pre- and post-inoculation radiation exposure in white mice inoculated with Swiss influenza virus in data of Dubin *et al.* [8], evaluated using a Cox proportional hazards model**

| Model | Dose post-inoculation ERR / Gy (+95% CI) | $p$-value |
|---|---|---|
| Linear | -0.13 (-0.35, 0.27) | 0.451 |
| Loglinear | -0.14 (-0.51, 0.23) | 0.454 |
| Model | Dose pre-inoculation ERR / Gy (+95% CI) | $p$-value |
| Linear | -0.62 (-0.90, -0.09) | 0.029 |
| Loglinear | -0.92 (-1.84, -0.10) | 0.028 |

**Table B8. Risks of death in mice associated with radiation exposure post inoculation to FA strain mouse encephalitis virus in the data of Tanner and McConchie [25], via fit of a logistic model**

| Model type | Adjustment | EOR / Gy + 95% CI | $p$-value |
|---|---|---|---|
| Linear logistic | Unadjusted | 0.09 (-0.09, 0.61) | 0.469 |
| Loglinear logistic | Unadjusted | 0.07 (-0.13, 0.27) | 0.490 |
| Linear logistic | Adjusted for ln[virus concentration] | 0.08 (-0.10, 0.59) | 0.525 |
| Loglinear logistic | Adjusted for ln[virus concentration] | 0.06 (-0.14, 0.26) | 0.544 |

**Table B9. Mortality and morbidity risks in Germantown white mice associated with pre-inoculation radiation exposure to mouse-adapted or egg-adapted PR8 influenza A virus in the data of Beutler and Gezon [26], via fit of a logistic model**

| Type of virus adaptation | Model | Adjustment | EOR / Gy + 95% CI | *p*-value |
|---|---|---|---|---|
| | | Mortality | | |
| Mouse-adapted | Linear logistic | Unadjusted | 0.16 (0.06, 0.30) | <0.001 |
| Mouse-adapted | Linear logistic | Adjusted for virus dilution | 0.23 (0.08, 0.43) | <0.001 |
| Mouse-adapted | Loglinear logistic | Unadjusted | 0.11 (0.04, 0.18) | 0.002 |
| Mouse-adapted | Loglinear logistic | Adjusted for virus dilution | 0.14 (0.06, 0.22) | <0.001 |
| Egg-adapted | Linear logistic | Unadjusted | 3.82 (1.13, 17.15) | <0.001 |
| Egg-adapted | Linear logistic | Adjusted for virus dilution | 4.14 (1.17, 19.06) | <0.001 |
| Egg-adapted | Loglinear logistic | Unadjusted | 0.84 (0.57, 1.14) | <0.001 |
| Egg-adapted | Loglinear logistic | Adjusted for virus dilution | 0.88 (0.57, 1.23) | <0.001 |
| | | Morbidity | | |
| Mouse-adapted | Linear logistic | Unadjusted | -0.17 (-0.18, -0.15) | <0.001 |
| Mouse-adapted | Linear logistic | Adjusted for virus dilution | -0.19 (-0.19, -0.18) | <0.001 |
| Mouse-adapted | Loglinear logistic | Unadjusted | -0.22 (-0.31, -0.13) | <0.001 |
| Mouse-adapted | Loglinear logistic | Adjusted for virus dilution | -0.34 (-0.45, -0.23) | <0.001 |
| Egg-adapted | Linear logistic | Unadjusted | 0.41 (0.14, 0.85) | <0.001 |
| Egg-adapted | Linear logistic | Adjusted for virus dilution | 1.02 (0.39, 2.17) | <0.001 |
| Egg-adapted | Loglinear logistic | Unadjusted | 0.22 (0.08, 0.36) | 0.001 |
| Egg-adapted | Loglinear logistic | Adjusted for virus dilution | 0.35 (0.16, 0.54) | <0.001 |

**Table B10. Risks of death in Swiss mice associated with pre-inoculation radiation exposure to *Pneumococcus* type III bacterial infection in the data of Hale and Stoner [27], via fit of a logistic model**

| Model | Subset of data used | EOR / Gy + 95% CI | $p$-value |
|---|---|---|---|
| Linear logistic | All animals | 0.09 (-0.04, 0.35) | 0.239 |
| Loglinear logistic | All animals | 0.07 (-0.05, 0.19) | 0.239 |
| Linear logistic | Only animals with intra-abdominal *pneumococcus* type III administered | 1.40 (0.39, 5.47) | <0.001 |
| Loglinear logistic | Only animals with intra-abdominal *pneumococcus* type III administered | 0.39 (0.21, 0.61) | <0.001 |

**Table B11. Risks of death in Swiss mice associated with pre-inoculation radiation exposure to mixture of influenza virus, *Pneumococcus* type III bacterial infection and larval infection by *Trichinella spiralis* in the data of Hale and Stoner [28], via fit of a linear logistic model**

|  | EOR / Gy + 95% CI | *p*-value |
|---|---|---|
| Unadjusted, all data | 0.11 (0.03, 0.21) | 0.004 |
| Unadjusted, fitted to experiments with influenza and *Pneumococcus* challenge infections only | 0.27 (0.13, 0.48) | <0.001 |
| Adjusted for type of challenge infection type[a] | 1.91 (1.12, 3.33) | <0.001 |
| Adjusted for challenge infection type, fitted to experiments with influenza and *Pneumococcus* challenge infections only[a] | 1.71 (0.97, 3.02) | <0.001 |

[a]influenza virus, *Pneumococcus* type III bacterial infection, larval infection by *Trichinella spiralis*

**Table B12. Risks of death in male C57BL mice associated with $^{60}$Co radiation exposure pre-inoculation to intranasally and intraperitoneally administered PR8 strain of type A influenza virus in the data of Quilligan *et al* [19], via fit of a linear logistic model**

|  | EOR / Gy + 95% CI | *p*-value |
|---|---|---|
| Assuming 6 mice in first control group | 3.73 (0.42, 85.85) | <0.001 |
| Assuming 8 mice in first control group | 4.24 (0.49, 96.97) | <0.001 |
| Assuming 10 mice in first control group | 4.75 (0.56, 108.10) | <0.001 |

**Table B13. Morbidity and mortality risks associated with pre-inoculation radiation exposure in CF-1 adult albino male mice inoculated with CAM A-prime strain of influenza virus in data of Berlin [9] via fit of a linear logistic model**

|  | EOR / Gy + 95% CI | *p*-value |
|---|---|---|
| Pneumonitis morbidity | -0.24 (-0.28, -0.17) | <0.001 |
| Pneumonitis mortality | -0.21 (-0.26, -0.14) | <0.001 |

**Table B14. Influenza A/PR8 morbidity and mortality risks associated with pre-inoculation radiation exposure in CF-1 adult albino male mice inoculated with CAM A-prime strain of influenza virus in data of Berlin and Cochran [10], evaluated using a linear logistic model**

|  | EOR / Gy + 95% CI | $p$-value |
|---|---|---|
| Table II data – analysis of influenza mortality and morbidity | | |
| Influenza morbidity | -0.11 (-0.18, 0.05) | 0.132 |
| Influenza mortality | -0.04 (-0.15, 0.26) | 0.711 |
| Table III data – analysis of influenza mortality considering mode of administration of virus | | |
| Unadjusted for mode of administration of influenza virus | 0.21 (0.03, 0.48) | 0.020 |
| Adjusted for mode of administration of influenza virus | 0.25 (0.05, 0.57) | 0.009 |

**Table B15. Mortality risks associated with pre-inoculation $^{144}$CeO$_2$ radiation exposure in C57BL/6J mouse inoculated with type A$_0$ influenza virus in data of Lundgren *et al* [29], evaluated using a linear logistic model**

|  | EOR / Gy + 95% CI | *p*-value |
|---|---|---|
| Mortality | 0.007 (0.002, 0.017) | 0.002 |

# References

1. Kirkby C, Mackenzie M. Is low dose radiation therapy a potential treatment for COVID-19 pneumonia? Radiother Oncol. 2020;147:221. doi: 10.1016/j.radonc.2020.04.004.
2. Ghadimi-Moghadam A, Haghani M, Bevelacqua JJ, Jafarzadeh A, Kaveh-Ahangar A, Mortazavi SMJ, et al. COVID-19 Tragic Pandemic: Concerns over Unintentional "Directed Accelerated Evolution" of Novel Coronavirus (SARS-CoV-2) and Introducing a Modified Treatment Method for ARDS. J Biomed Phys Engineering. 2020;10(2):241-6. doi: 10.31661/jbpe.v0i0.2003-1085.
3. ClinicalTrials.gov. ClinicalTrials.gov 18 studies found for COVID-19 Radiation https://clinicaltrials.gov/ct2/results?cond=COVID-19+Radiation&term=&cntry=&state=&city=&dist= 2020 [updated 8/2020; cited 2020 8/2020]. Available from: https://clinicaltrials.gov/ct2/results?cond=COVID-19+Radiation&term=&cntry=&state=&city=&dist=.
4. Calabrese EJ, Dhawan G. How radiotherapy was historically used to treat pneumonia: could it be useful today? Yale J Biol Med. 2013;86(4):555-70. Epub 2013/12/19. PubMed PMID: 24348219; PubMed Central PMCID: PMCPMC3848110.
5. Fried C. The Roentgen treatment of experimental pneumonia in the guinea-pig. Radiology. 1941;37(2):197-202. doi: 10.1148/37.2.197.
6. Lieberman LM, Hodes PJ, Leopold SS. Roentgen therapy of experimental lobar pneumonia in dogs. Am J Med Sciences. 1941;291(1):92-100.
7. Baylin GJ, Dubin IN, Gobbel WG, Jr. The effect of roentgen therapy on experimental virus pneumonia; on feline virus pneumonia. Am J Roentgenol Radium Ther. 1946;55:473-7. Epub 1946/04/01. PubMed PMID: 21024568.
8. Dubin IN, Baylin GJ, Gobble WG, Jr. The effect of roentgen therapy on experimental virus pneumonia; on pneumonia produced in white mice by swine influenza virus. Am J Roentgenol Radium Ther. 1946;55:478-81. Epub 1946/04/01. PubMed PMID: 21024569.
9. Berlin BS. Sparing Effect of X-Rays for Mice Inoculated Intranasally with Egg-Adapted Influenza Virus, Cam Strain. Proceedings of the Society for Experimental Biology and Medicine Society for Experimental Biology and Medicine. 1964;117:864-9. Epub 1964/12/01. doi: 10.3181/00379727-117-29720. PubMed PMID: 14245857.
10. Berlin BS, Cochran KW. Delay of fatal pneumonia in x-irradiated mice inoculated with mouse-adapted influenza virus, PR8 strain. Radiat Res. 1967;31(2):343-51. Epub 1967/06/01. PubMed PMID: 6025867.
11. Taliaferro WH, Taliaferro LG. Effect of x-rays on immunity; a review. J Immunol. 1951;66(2):181-212. Epub 1951/02/01. PubMed PMID: 14814301.
12. Wikipedia. Roentgen (unit) (https://en.wikipedia.org/wiki/Roentgen_(unit)) Wikipedia; 2020 [cited 2020 4/2020]. Available from: https://en.wikipedia.org/wiki/Roentgen_(unit).
13. Murphy JB, Sturm E. A comparison of the effects of X-Ray and dry heat on antibody formation. J Exp Med. 1925;41(2):245-55. Epub 1925/01/31. doi: 10.1084/jem.41.2.245. PubMed PMID: 19868985; PubMed Central PMCID: PMCPMC2130939.
14. Cox DR. Regression models and life-tables. J Royal Statist Soc Series B. 1972;34(2):187-220.
15. Fried C. Die artefizielle Pneumonie und lhre Bestrahlung. Experimentalbeitrag zur Frage der Wirkung der Röntgenstrahlen auf Entzündungsgewebe. Strahlentherapie. 1937;58:430-48.
16. Cytel I. LogXact 11 version 11.0.0. Cambridge, MA 02139: Cytel, Inc.; 2015.
17. Breslow NE, Day NE. Statistical methods in cancer research. Volume II--The design and analysis of cohort studies. IARC SciPubl. 1987;(82):1-406.
18. McCullagh P, Nelder JA. Generalized linear models. 2nd edition. Boca Raton, FL: Chapman and Hall/CRC; 1989 1989. 1-526 p.
19. Quilligan JJ, Jr., Boche RD, Carruthers EJ, Axtell SL, Trivedi JC. Continuous Cobalt-60 Irradiation and Immunity to Influenza Virus. J Immunol. 1963;90:506-11. Epub 1963/04/01. PubMed PMID: 14082011.

20. McCullagh P. Regression models for ordinal data. J Royal Statist Soc Series B (Methodological). 1980;42(2):109-42.
21. Efron B. Better bootstrap confidence intervals. J Am Statist Assoc. 1987;82(397):171-85. doi: 10.2307/2289144.
22. Risk Sciences International. Epicure version 2.0.1.0. 2.0.1.0 ed. 55 Metcalfe, K1P 6L5, Canada: Risk Sciences International; 2015.
23. R Project version 3.6.1. R: A language and environment for statistical computing. version 3.6.1 https://www.r-project.org. 3.6.1 ed. Vienna, Austria: R Foundation for Statistical Computing; 2019.
24. Bond VP, Shechmeister IL, Swift MN, Fishler MC. The effects of X irradiation on a naturally occurring endemic infection. J Infect Dis. 1952;91(1):26-32. Epub 1952/07/01. doi: 10.1093/infdis/91.1.26. PubMed PMID: 14946423.
25. Tanner WA, McConchie JE. Effect of roentgen therapy on mouse encephalitis. Radiology. 1949;53(1):101-3. Epub 1949/07/01. doi: 10.1148/53.1.101. PubMed PMID: 18131939.
26. Beutler E, Gezon HM. The effect of total body X irradiation on the susceptibility of mice to influenza A virus infection. J Immunol. 1952;68(3):227-42.
27. Hale WM, Stoner RD. The effect of cobalt-60 gamma radiation on passive immunity. Yale J Biol Med. 1953;25(5):326-33. Epub 1953/04/01. PubMed PMID: 13057263; PubMed Central PMCID: PMCPMC2599518.
28. Hale WM, Stoner RD. Effects of ionizing radiation on immunity. Radiat Res. 1954;1(5):459-69. Epub 1954/10/01. PubMed PMID: 13204561.
29. Lundgren DL, Sanchez A, Thomas RL, Chiffelle TL, McClellan RO. Effects of inhaled 144CeO2 on influenza virus infection in mice. Proceedings of the Society for Experimental Biology and Medicine Society for Experimental Biology and Medicine. 1973;144(1):238-44. Epub 1973/10/01. doi: 10.3181/00379727-144-37564. PubMed PMID: 4771564.
30. Hasegawa H, Kadowaki S, Takahashi H, Iwasaki T, Tamura S, Kurata T. Protection against influenza virus infection by nasal vaccination in advance of sublethal irradiation. Vaccine. 2000;18(23):2560-5. Epub 2000/04/25. doi: 10.1016/s0264-410x(99)00553-8. PubMed PMID: 10775790.
31. Dadachova E, Burns T, Bryan RA, Apostolidis C, Brechbiel MW, Nosanchuk JD, et al. Feasibility of radioimmunotherapy of experimental pneumococcal infection. Antimicrob Agents Chemother. 2004;48(5):1624-9. doi: 10.1128/aac.48.5.1624-1629.2004.
32. Rödel F, Arenas M, Ott OJ, Fournier C, Georgakilas AG, Tapio S, et al. Low-dose radiation therapy for COVID-19 pneumopathy: what is the evidence? Strahlenther Onkol. 2020;196(8):679-82. Epub 2020/05/11. doi: 10.1007/s00066-020-01635-7. PubMed PMID: 32388805; PubMed Central PMCID: PMCPMC7211051.
33. Schaue D, McBride WH. Flying by the seat of our pants: is low dose radiation therapy for COVID-19 an option? Int J Radiat Biol. 2020:1-5. doi: 10.1080/09553002.2020.1767314.
34. Kirsch DG, Diehn M, Cucinotta FA, Weichselbaum R. Lack of supporting data make the risks of a clinical trial of radiation therapy as a treatment for COVID-19 pneumonia unacceptable. Radiother Oncol. 2020;147:217-20. doi: https://doi.org/10.1016/j.radonc.2020.04.060.
35. United Nations Scientific Committee on the Effects of Atomic Radiation (UNSCEAR). UNSCEAR 2006 Report. Annex A. Epidemiological Studies of Radiation and Cancer. New York: United Nations; 2008 2008. 13-322 p.
36. Little MP, Azizova TV, Bazyka D, Bouffler SD, Cardis E, Chekin S, et al. Systematic review and meta-analysis of circulatory disease from exposure to low-level ionizing radiation and estimates of potential population mortality risks. Environ Health Perspect. 2012;120(11):1503-11. doi: 10.1289/ehp.1204982 [doi].